# Hardware-conscious Software Training for Deep Neural Network Inference Accelerator Chips to Recover Accuracy Degradation due to Hardware Variabilities

Shuchao Gao and Takashi Ohsawa

[1] Graduate School of Information, Production and Systems, Waseda University
Hibikino 2-7, Wakamatsu-ku, Kitakyushu, Fukuoka 808-0135, Japan
Phone: +81-93-692-5349, E-mail: takashi.ohsawa@aoni.waseda.jp

**Abstract**

**Deep neural network (DNN) has been widely applied in various industries. Specialized chips are being discussed for the purpose of achieving lower power consumption with higher throughput. Hardware variations introduced during the process of chip manufacturing are the main reason for affecting the inference accuracies. In this paper, we propose hardware-conscious software training (HCST) method which enables high inference accuracies even under the influence of hardware variations.**

## 1. Introduction

The accuracies of deep neural network (DNN) inference accelerators are lower than the corresponding accuracies in software if the accelerators are trained offline because of the inevitable hardware variabilities. To overcome this, several methods have been proposed such as the on-chip training [1] and the in-situ training [2]. However, they suffer from endurance limitation of the nonvolatile memory devices which are used for storing the weights, because they are required to perform a huge number of device switchings after training iterations (epochs) in these trainings [3]. We propose a software training in which the hardware variabilities such as offset voltages in operational amplifiers (op-amps) are taken into consideration so that the accuracy degradation due to the hardware parameters' fluctuations are compensated. The method makes the memory devices free from the endurance constraint, the nonlinearity and the asymmetry issues in updating the resistances.

## 2. Hardware-conscious software training (HCST)

The proposed method named hardware-conscious software training (HCST) is performed by a training program written by using parameters of electronic circuits such as voltage, current and conductance. It, furthermore, embodies finite open loop gains (G's) and offset voltages ($V_{os}$'s) of op-amps used in DNN inference accelerators which are considered as hardware imperfections. $V_{os}$'s in the program are set to actually fluctuated values which are measured from each hardware chip, making the training individual for each chip.

Fig. 1 shows an $l$-th synapse crossbar array in which each weight is represented by a pair of ReRAM devices ($g_{ji}^{(l)+}$, $g_{ji}^{(l)-}$) which are connected to a pair of bit-lines ($BL_j^{(l)+}$, $BL_j^{(l)-}$) and the currents flowing through the bit-lines are subtracted after converted to voltages for realizing a positive or a negative weight. The circuit diagrams of the IV-converter (IVC), subtractor, ReLU, and voltage follower are shown in Fig. 2 (a)-(d) and the characteristic curve of ReLU is shown in Fig. 2 (e). The op-amps have offset voltages $V_{os}$'s and finite open loop gains G's except for ReLU. The op-amps used in the circuits have the same structure (Fig. 3) of Type 1 and Type 2 with different sizes in 32nm technology. Type 1 is used in IVC, subtractor, and ReLU. Type 2 is used in voltage follower. $V_{th}$ fluctuation's standard deviation in each MOSFET is estimated by using the equation proposed in [4]. Fig. 4 (a) and (b) respectively show Monte Carlo simulation results of $V_{os}$'s for Type 1 and Type 2 based on the $V_{th}$ fluctuations in Tabel I and II, respectively. In this study, we assigned $V_{os}$'s of each op-amp based on this standard deviation chip by chip, though they actually are to be measured from each fabricated chip. The process of HCST is illustrated in Fig. 5. The training completes in the training program (hardware emulator (HWE)) with the measured (but assigned in this study) $V_{os}$'s to set $g_{ji}$ to their final values $g_{ji}^{(f)}$. We conduct HCST as an individual training for each chip with different $V_{os}$'s (Fig. 6).

## 3. Training program including hardware variations

Fig. 7 and Fig. 8 illustrate the formulae for forward and backward propagations in the HWE where the unit errors are derived in the backward path according to the chain rule in the neuron circuits. In the both propagations, each op-amp's G and $V_{os}$ are taken into account. The updated conductance is deployed to the both $g_{ji}^{(l)\pm}$ in the synapse array according to their sign as shown in Fig. 9. Depending on the restriction of ReRAM, $g_{ji}^{(l)\pm}=g_{max}$ when $g_{ji}^{(l)\pm} > g_{max}$ and $g_{ji}^{(l)\pm}=0$ when $g_{ji}^{(l)\pm} < g_{min}$ as shown in Fig. 9. The ReRAM without electroforming can be used as the 0 conductance device.

## 4. Application to IRIS dataset

The outputs simulated by HSPICE for three kinds of flowers in the IRIS dataset [5] are shown in Fig. 10 without $V_{os}$'s. The network has $4 \times 5 \times 3$ struture. The training result of the HWE without $V_{os}$'s is 97.33% (baseline) which is the same as the HSPICE results. Fig. 11 (a) and (b) show 10 chips' accuracies of the baseline (black) without $V_{os}$'s, offline training (blue) and HCST (red) both with $V_{os}$'s for HWE and HSPICE, respectively.

## 5. Conclusion

It was verified by HSPICE that the HCST is effective for recovering the accuracy degradation due to the hardware variabilities in DNN inference accelerators.

**Acknowledgements**

This work was supported by VLSI Design and Education Center (VDEC), the University of Tokyo with the collaboration with Synopsys Corporation.

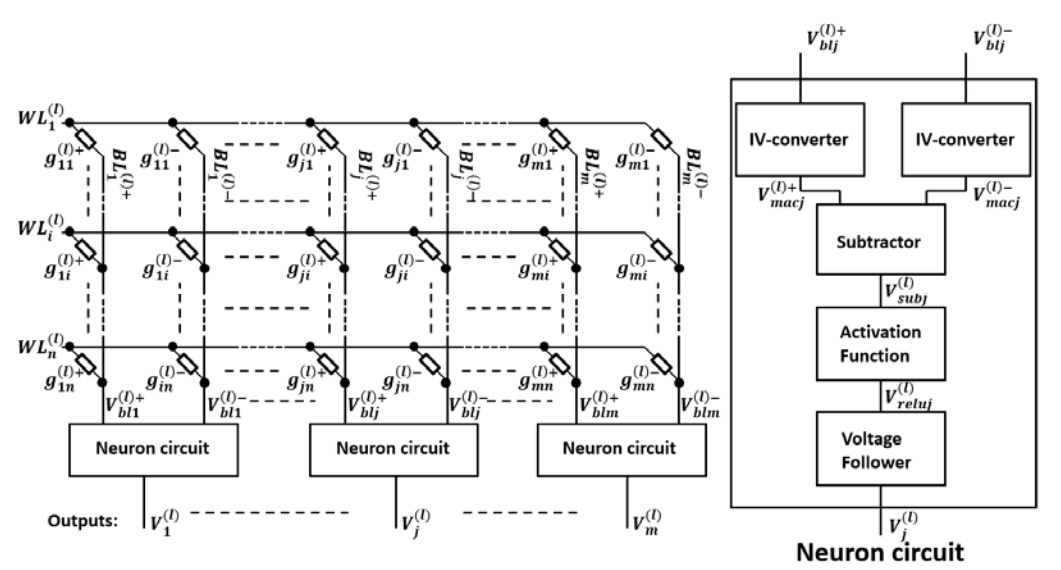

Fig. 1 $l$-th synapse array and neuron circuits in a DNN inference accelerator.

Fig. 2 Circuits of (a) IV-converter, (b) voltage subtractor, (c) ReLU and (d) voltage follower with offset voltages. (e) The characteristic curve of the ReLU

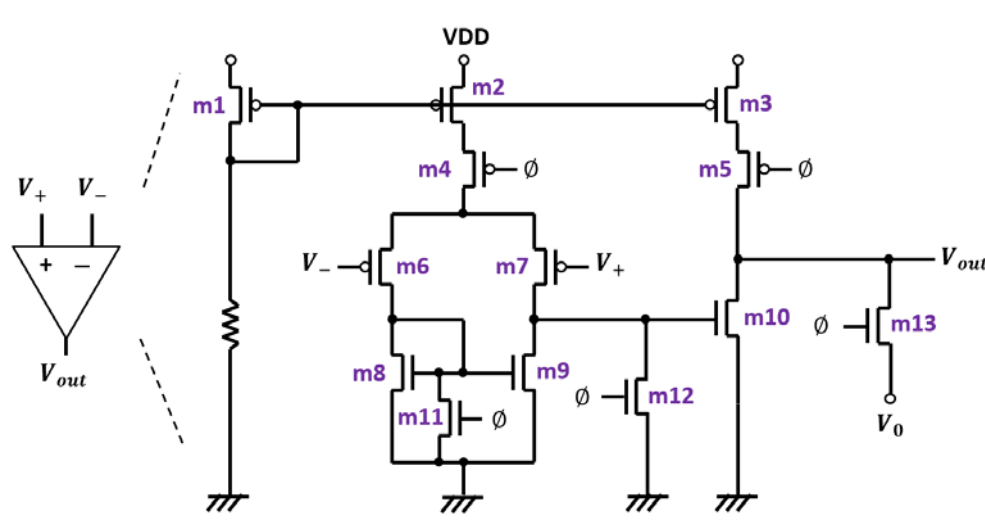

Fig. 3 Circuit schematic of the op-amp

Table I MOSFETs in op-amp Type 1 with G=1515.

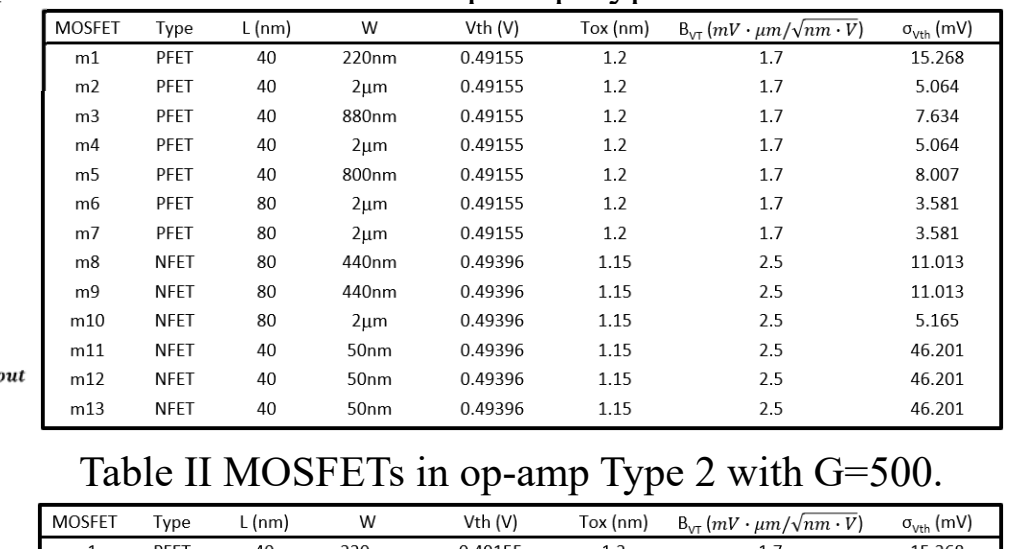

| MOSFET | Type | L (nm) | W | Vth (V) | Tox (nm) | $B_{VT}$ $(mV \cdot \mu m/\sqrt{nm \cdot V})$ | $\sigma_{Vth}$ (mV) |
|---|---|---|---|---|---|---|---|
| m1 | PFET | 40 | 220nm | 0.49155 | 1.2 | 1.7 | 15.268 |
| m2 | PFET | 40 | 2μm | 0.49155 | 1.2 | 1.7 | 5.064 |
| m3 | PFET | 40 | 880nm | 0.49155 | 1.2 | 1.7 | 7.634 |
| m4 | PFET | 40 | 2μm | 0.49155 | 1.2 | 1.7 | 5.064 |
| m5 | PFET | 40 | 800nm | 0.49155 | 1.2 | 1.7 | 8.007 |
| m6 | PFET | 80 | 2μm | 0.49155 | 1.2 | 1.7 | 3.581 |
| m7 | PFET | 80 | 2μm | 0.49155 | 1.2 | 1.7 | 3.581 |
| m8 | NFET | 80 | 440nm | 0.49396 | 1.15 | 2.5 | 11.013 |
| m9 | NFET | 80 | 440nm | 0.49396 | 1.15 | 2.5 | 11.013 |
| m10 | NFET | 80 | 2μm | 0.49396 | 1.15 | 2.5 | 5.165 |
| m11 | NFET | 40 | 50nm | 0.49396 | 1.15 | 2.5 | 46.201 |
| m12 | NFET | 40 | 50nm | 0.49396 | 1.15 | 2.5 | 46.201 |
| m13 | NFET | 40 | 50nm | 0.49396 | 1.15 | 2.5 | 46.201 |

Table II MOSFETs in op-amp Type 2 with G=500.

| MOSFET | Type | L (nm) | W | Vth (V) | Tox (nm) | $B_{VT}$ $(mV \cdot \mu m/\sqrt{nm \cdot V})$ | $\sigma_{Vth}$ (mV) |
|---|---|---|---|---|---|---|---|
| m1 | PFET | 40 | 220nm | 0.49155 | 1.2 | 1.7 | 15.268 |
| m2 | PFET | 40 | 40μm | 0.49155 | 1.2 | 1.7 | 1.132 |
| m3 | PFET | 40 | 400μm | 0.49155 | 1.2 | 1.7 | 0.358 |
| m4 | PFET | 40 | 40μm | 0.49155 | 1.2 | 1.7 | 1.132 |
| m5 | PFET | 40 | 400μm | 0.49155 | 1.2 | 1.7 | 0.358 |
| m6 | PFET | 80 | 40μm | 0.49155 | 1.2 | 1.7 | 0.801 |
| m7 | PFET | 80 | 40μm | 0.49155 | 1.2 | 1.7 | 0.801 |
| m8 | NFET | 80 | 8.8μm | 0.49396 | 1.15 | 2.5 | 2.463 |
| m9 | NFET | 80 | 8.8μm | 0.49396 | 1.15 | 2.5 | 2.463 |
| m10 | NFET | 80 | 200μm | 0.49396 | 1.15 | 2.5 | 0.517 |
| m11 | NFET | 40 | 50nm | 0.49396 | 1.15 | 2.5 | 46.201 |
| m12 | NFET | 40 | 50nm | 0.49396 | 1.15 | 2.5 | 46.201 |
| m13 | NFET | 40 | 50nm | 0.49396 | 1.15 | 2.5 | 46.201 |

Fig. 4 Monte Carlo simulation results for the off-set voltages of the op-amp (a) Type 1 and (b) Type 2.

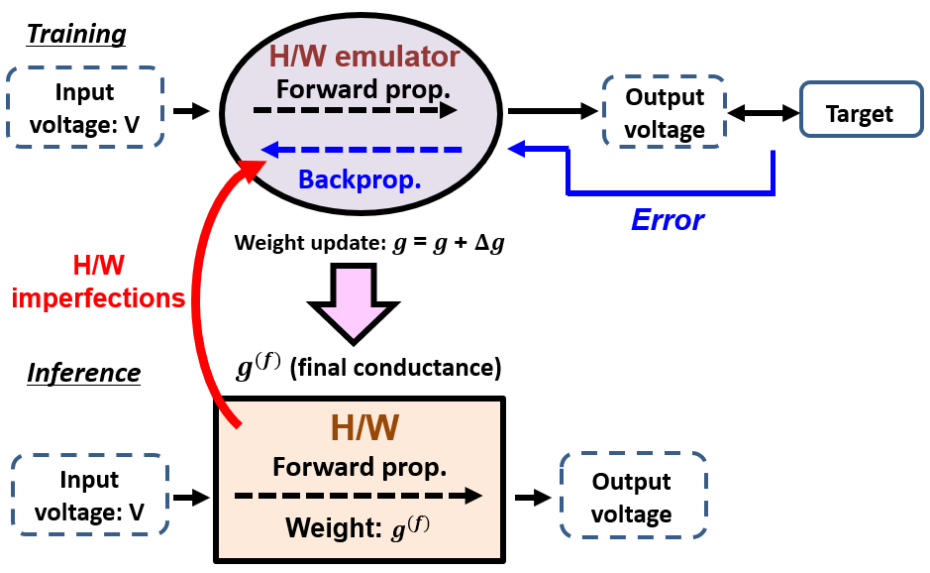

Fig. 5 Concept of hardware (H/W)-conscious software training (HCST)

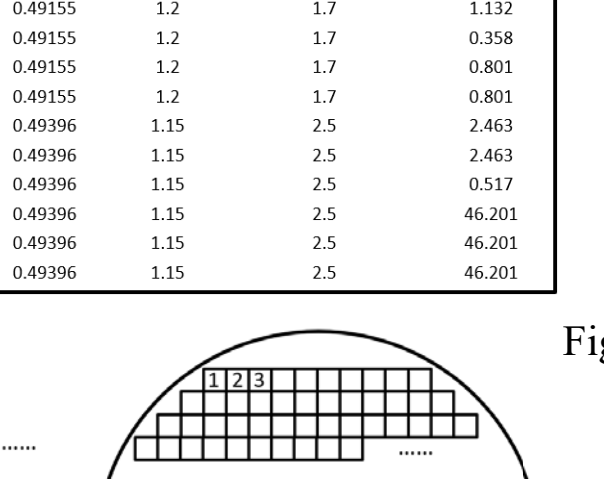

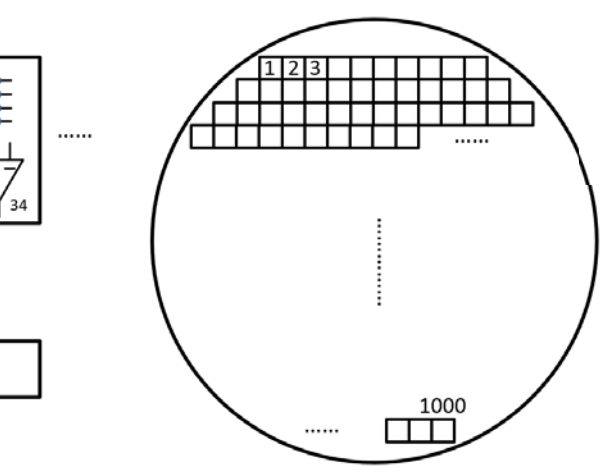

Fig. 6 The method of HCST as an individual training for each chip

IF $g_{ji}^{(l)+} > 0$:
THEN: $g_{ji}^{(l)+} = g_{ji}^{(l)+} + \Delta g_{ji}^{(l)+}$
IF $g_{ji}^{(l)+} < 0$:
THEN: $g_{ji}^{(l)-} = -g_{ji}^{(l)+}$ , $g_{ji}^{(l)+} = 0$
ElSE IF $g_{ji}^{(l)-} > 0$:
THEN: $g_{ji}^{(l)-} = g_{ji}^{(l)-} + \Delta g_{ji}^{(l)-}$
IF $g_{ji}^{(l)-} < 0$:
THEN: $g_{ji}^{(l)+} = -g_{ji}^{(l)-}$ , $g_{ji}^{(l)-} = 0$
IF $g_{ji}^{(l)\pm} > g_{max}$ :
THEN: $g_{ji}^{(l)\pm} = g_{max}$
ElSE IF $g_{ji}^{(l)\pm} < g_{min}$ :
THEN: $g_{ji}^{(l)\pm} = 0$

Fig. 9 Update algorithm

**IV-Converter:**

$$\sum_{i=1}^{n} g_{ji}^{(l)+}(V_i^{(l-1)} - V_{blj}^{(l)+}) = g_L\,(V_{blj}^{(l)+} - V_{macj}^{(l)+}) \qquad V_{macj}^{(l)+} - V_{mac0} = G_I\,\left[\left(V_0 + V_{osIj}^{(l)+}\right) - V_{blj}^{(l)+}\right]$$

$$\Rightarrow\quad V_{macj}^{(l)+} = \frac{\left(\sum_{i=1}^{n} g_{ji}^{(l)+} + g_L\right)V_{mac0} + G_I\,\left[\left(V_0 + V_{osIj}^{(l)+}\right)\left(\sum_{i=1}^{n} g_{ji}^{(l)+} + g_L\right) - \sum_{i=1}^{n} g_{ji}^{(l)+} V_i^{(l-1)}\right]}{G_I\,g_L + \sum_{i=1}^{n} g_{ji}^{(l)+} + g_L}$$

$$\sum_{i=1}^{n} g_{ji}^{(l)-}(V_i^{(l-1)} - V_{blj}^{(l)-}) = g_L\,(V_{blj}^{(l)-} - V_{macj}^{(l)-}) \qquad V_{macj}^{(l)-} - V_{mac0} = G_I\,\left[\left(V_0 + V_{osIj}^{(l)-}\right) - V_{blj}^{(l)-}\right]$$

$$\Rightarrow\quad V_{macj}^{(l)-} = \frac{\left(\sum_{i=1}^{n} g_{ji}^{(l)-} + g_L\right)V_{mac0} + G_I\,\left[\left(V_0 + V_{osIj}^{(l)-}\right)\left(\sum_{i=1}^{n} g_{ji}^{(l)-} + g_L\right) - \sum_{i=1}^{n} g_{ji}^{(l)-} V_i^{(l-1)}\right]}{G_I\,g_L + \sum_{i=1}^{n} g_{ji}^{(l)-} + g_L}$$

**Subtractor:**

$$V_{subj}^{(l)} - V_{sub0} = G_S\,\left[\left(\frac{V_0 + V_{macj}^{(l)-}}{2} + V_{osSj}^{(l)}\right) - \frac{V_{macj}^{(l)+} + V_{subj}^{(l)}}{2}\right]$$

**ReLU:**

$$\begin{cases} V_{reluj}^{(l)} = V_{subj}^{(l)} & (V_{subj}^{(l)} > V_0 - V_{osRj}^{(l)}) \\ V_{reluj}^{(l)} = V_0 & (V_{subj}^{(l)} \le V_0 - V_{osRj}^{(l)}) \end{cases}$$

**Voltage Follower:**

$$V_j^{(l)} - V_{F0} = G_F\,\left[\left(V_{reluj}^{(l)} + V_{osFj}^{(l)}\right) - V_j^{(l)}\right]$$

Fig. 7 Formula for forward propagation

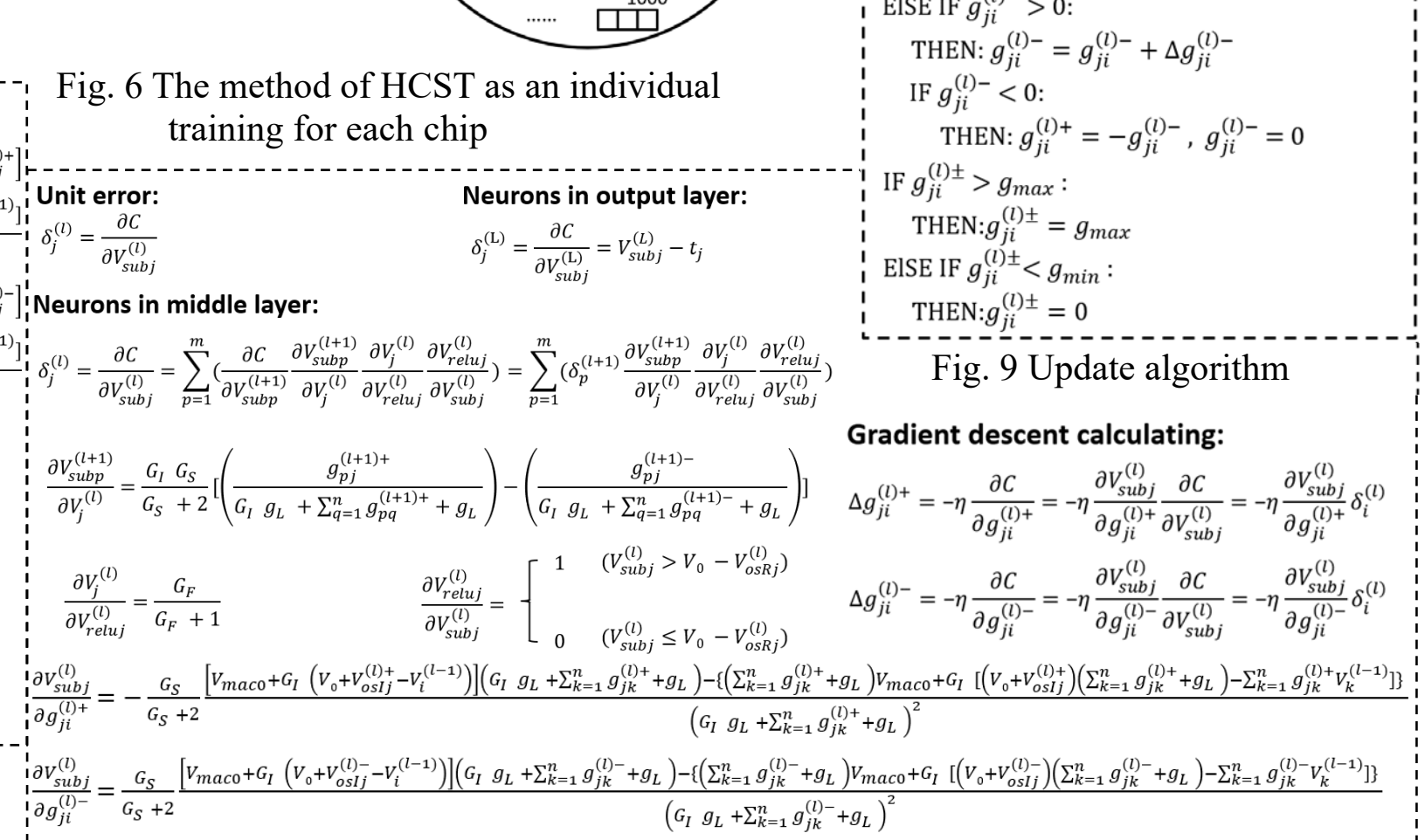

**Unit error:**

$$\delta_j^{(l)} = \frac{\partial C}{\partial V_{subj}^{(l)}}$$

**Neurons in output layer:**

$$\delta_j^{(L)} = \frac{\partial C}{\partial V_{subj}^{(L)}} = V_{subj}^{(L)} - t_j$$

**Neurons in middle layer:**

$$\delta_j^{(l)} = \frac{\partial C}{\partial V_{subj}^{(l)}} = \sum_{p=1}^{m}\left(\frac{\partial C}{\partial V_{subp}^{(l+1)}}\frac{\partial V_{subp}^{(l+1)}}{\partial V_j^{(l)}}\frac{\partial V_j^{(l)}}{\partial V_{reluj}^{(l)}}\frac{\partial V_{reluj}^{(l)}}{\partial V_{subj}^{(l)}}\right) = \sum_{p=1}^{m}\left(\delta_p^{(l+1)}\frac{\partial V_{subp}^{(l+1)}}{\partial V_j^{(l)}}\frac{\partial V_j^{(l)}}{\partial V_{reluj}^{(l)}}\frac{\partial V_{reluj}^{(l)}}{\partial V_{subj}^{(l)}}\right)$$

$$\frac{\partial V_{subp}^{(l+1)}}{\partial V_j^{(l)}} = \frac{G_I\;G_S}{G_S\;+2}\left[\left(\frac{g_{pj}^{(l+1)+}}{G_I\;g_L\;+\sum_{q=1}^{n} g_{pq}^{(l+1)+} + g_L}\right) - \left(\frac{g_{pj}^{(l+1)-}}{G_I\;g_L\;+\sum_{q=1}^{n} g_{pq}^{(l+1)-} + g_L}\right)\right]$$

$$\frac{\partial V_j^{(l)}}{\partial V_{reluj}^{(l)}} = \frac{G_F}{G_F\;+1} \qquad \frac{\partial V_{reluj}^{(l)}}{\partial V_{subj}^{(l)}} = \begin{cases} 1 & (V_{subj}^{(l)} > V_0 - V_{osRj}^{(l)}) \\ 0 & (V_{subj}^{(l)} \le V_0 - V_{osRj}^{(l)}) \end{cases}$$

$$\frac{\partial V_{subj}^{(l)}}{\partial g_{ji}^{(l)+}} = -\frac{G_S}{G_S\;+2}\,\frac{\left[V_{maco} + G_I\,\left(V_0 + V_{osIj}^{(l)+} - V_i^{(l-1)}\right)\right]\left(G_I\;g_L + \sum_{k=1}^{n} g_{jk}^{(l)+} + g_L\right) - \{\left(\sum_{k=1}^{n} g_{jk}^{(l)+} + g_L\right)V_{maco} + G_I\,\left[\left(V_0 + V_{osIj}^{(l)+}\right)\left(\sum_{k=1}^{n} g_{jk}^{(l)+} + g_L\right) - \sum_{k=1}^{n} g_{jk}^{(l)+} V_k^{(l-1)}\right]\}}{\left(G_I\;g_L + \sum_{k=1}^{n} g_{jk}^{(l)+} + g_L\right)^2}$$

$$\frac{\partial V_{subj}^{(l)}}{\partial g_{ji}^{(l)-}} = \frac{G_S}{G_S\;+2}\,\frac{\left[V_{maco} + G_I\,\left(V_0 + V_{osIj}^{(l)-} - V_i^{(l-1)}\right)\right]\left(G_I\;g_L + \sum_{k=1}^{n} g_{jk}^{(l)-} + g_L\right) - \{\left(\sum_{k=1}^{n} g_{jk}^{(l)-} + g_L\right)V_{maco} + G_I\,\left[\left(V_0 + V_{osIj}^{(l)-}\right)\left(\sum_{k=1}^{n} g_{jk}^{(l)-} + g_L\right) - \sum_{k=1}^{n} g_{jk}^{(l)-} V_k^{(l-1)}\right]\}}{\left(G_I\;g_L + \sum_{k=1}^{n} g_{jk}^{(l)-} + g_L\right)^2}$$

**Gradient descent calculating:**

$$\Delta g_{ji}^{(l)+} = -\eta\frac{\partial C}{\partial g_{ji}^{(l)+}} = -\eta\frac{\partial V_{subj}^{(l)}}{\partial g_{ji}^{(l)+}}\frac{\partial C}{\partial V_{subj}^{(l)}} = -\eta\frac{\partial V_{subj}^{(l)}}{\partial g_{ji}^{(l)+}}\delta_i^{(l)}$$

$$\Delta g_{ji}^{(l)-} = -\eta\frac{\partial C}{\partial g_{ji}^{(l)-}} = -\eta\frac{\partial V_{subj}^{(l)}}{\partial g_{ji}^{(l)-}}\frac{\partial C}{\partial V_{subj}^{(l)}} = -\eta\frac{\partial V_{subj}^{(l)}}{\partial g_{ji}^{(l)-}}\delta_i^{(l)}$$

Fig. 8 Formula for backward propagation

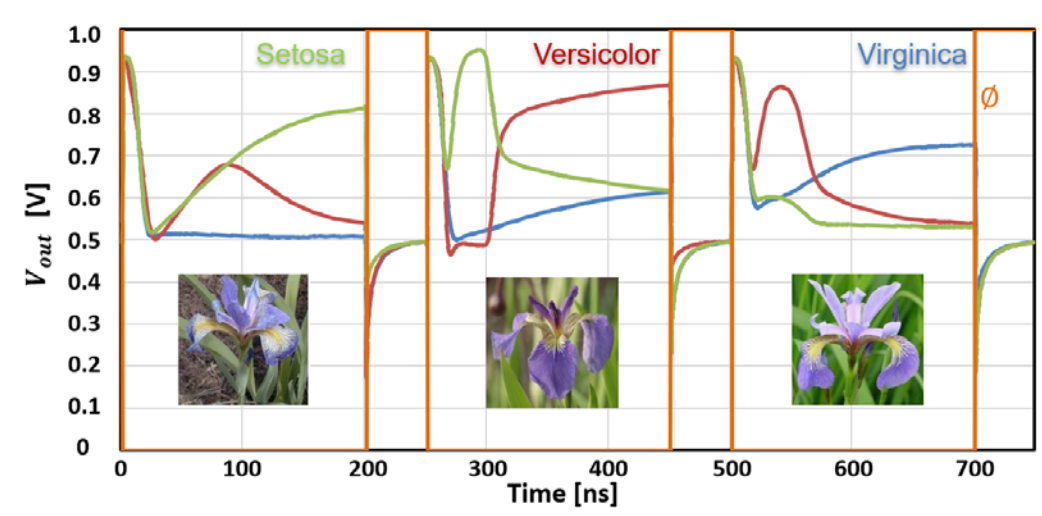

Fig. 10 Simulation waveforms of the three outputs (green, red and blue for Setosa, Versicolor and Virginica, respectively). The orange line is clock Ø.

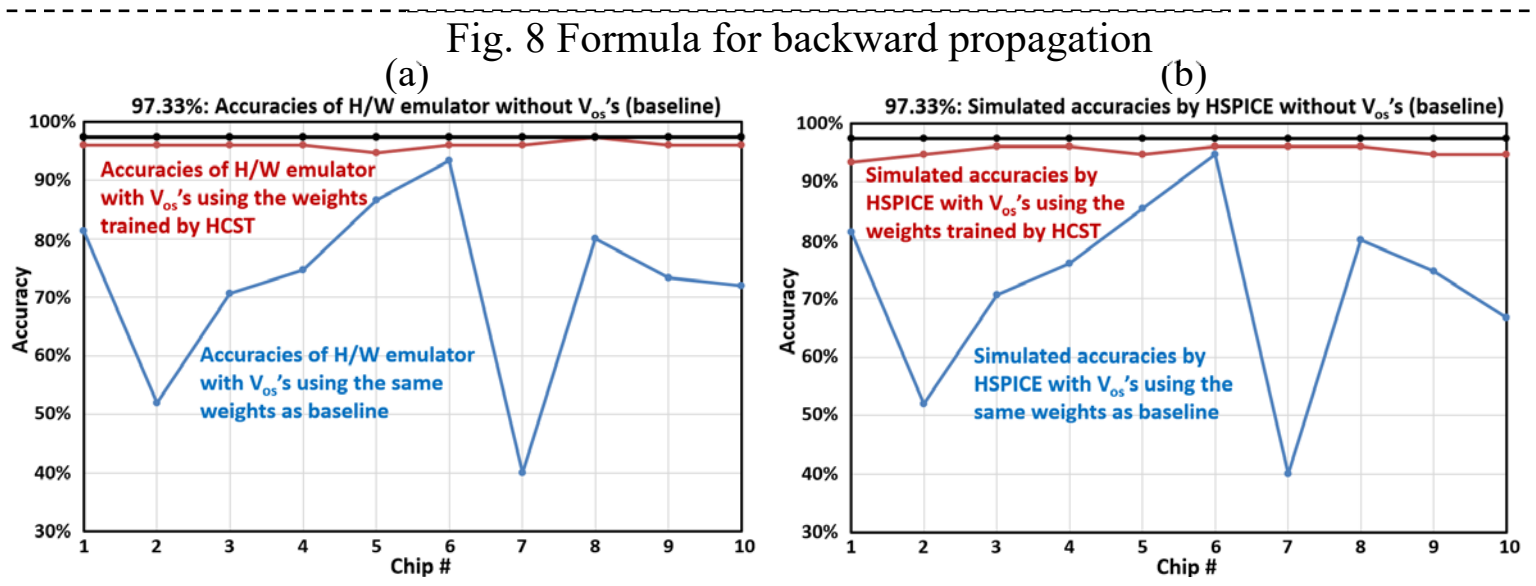

Fig. 11 (a) Results in H/W emulator and (b) results in HSPICE simulation